\documentclass[aps,prc,twocolumn,groupedaddress]{revtex4}
\usepackage{amsmath,amssymb}
\usepackage[dvipdfm]{graphicx,color}
\usepackage{subfigure}
\usepackage{longtable}

\begin{document}

%\preprint{}

\title{Analysis of the effect of core structure upon dineutron correlation using antisymmetrized molecular dynamics}


\author{Fumiharu Kobayashi}
%\thanks{}
\affiliation{Department of Physics, Niigata University, Niigata 950-2181, Japan}

\author{Yoshiko Kanada-En'yo}
%\thanks{}
\affiliation{Department of Physics, Kyoto University, Kyoto 606-8502, Japan}

\date{\today}

\begin{abstract}
We extend the method of antisymmetrized molecular dynamics to investigate dineutron correlation.
We apply this method to $^{10}$Be as an example and investigate the motion of two neutrons 
around a largely deformed $^8$Be core 
by analyzing the two-neutron overlap function around the core.
We show that the core structure plays an important role in 
dineutron formation and expansion from the core
and that the present framework is effective for the studies of dineutron correlation.
\end{abstract}

%\keywords{}

\maketitle

\section{Introduction}
\label{sec:introduction}

Many exotic phenomena have been found in the neutron-rich nuclei, 
and more have been suggested by both theoretical and experimental studies. 
One such phenomenon is dineutron correlation, 
a strong spatial correlation between two neutrons coupled to a spin singlet. 
Although the two neutrons are not bound in a free space, 
strong dineutron correlation has been theoretically suggested, 
e.g., in a low-density region of nuclear matter \cite{baldo90, matsuo06}
or in the neutron-halo or -skin regions of neutron-rich nuclei
\cite{bertsch91,zhukov93,arai01,descouvemont03,matsuo05,hagino05,
descouvemont06,enyo07,hagino07,pillet07,itagaki08,pillet10,
kobayashi12,kobayashi13,kobayashi14}. 
These studies clarify that the strength of dineutron correlation reflected in the dineutron size changes significantly 
depending on circumstances such as nuclear density 
and potential from the core. 
In addition, the dineutron and diproton correlations have been intensively discussed 
via $2n$ and $2p$ emissions from the unbound nuclei 
in connection with recent experiments
\cite{grigorenko00,grigorenko01,grigorenko03,grigorenko07,grigorenko09,grigorenko09_2,
johansson10,spyrou12,egorova12,grigorenko12,kohley13,kohley15}. 

Preceding studies have investigated dineutron correlation in the ground and excited states in certain nuclei, 
but they have not conducted a systematic investigation, 
and the formation mechanism underlying dineutron correlation 
and the dynamics of one or more dineutrons are not well understood. 
To investigate in detail the dineutron motion with respect to the core, 
a core$+2n$ three-body model is useful 
\cite{bertsch91,zhukov93,arai01,descouvemont03,hagino05,descouvemont06,hagino07}.
However, in three-body models, it is somewhat difficult to take various structure changes into account
(including excitation, deformation, and clustering), 
and inert, spherical cores have been assumed in most of the preceding studies. 
Core excitation and deformation can affect dineutron correlation, 
and it is necessary to consider changes in the core structure explicitly
for a systematic investigation of dineutron correlation. 

In our previous studies, using dineutron condensate (DC) wave functions
\cite{kobayashi12,kobayashi13,kobayashi14}, 
we showed that dineutrons in nuclei are fragile and easily broken. 
To study the properties of dineutron correlation, 
the following dineutron-breaking effects should be considered. 
First, the dissociation of a spin-singlet two-neutron pair 
due to the spin-orbit potential from the core,  
as discussed in Refs.~\cite{kobayashi14,enyo14,enyo15}. 
Without the spin-orbit potential, 
two neutrons tend to be coupled to a spin-singlet pair 
because of the spin-singlet $s$-wave attraction. 
However, in reality, 
two valence neutrons at the nuclear surface feel the spin-orbit potential from the core 
and tend to occupy the $LS$-favored orbits,  
resulting in mixing of the spin-triplet pair, 
i.e., reduction of the spin-singlet dineutron component. 
We call this effect at the surface due to the spin-orbit potential 
the ``$LS$ dissociation'' of a dineutron. 
Second, the swell in the size of a dineutron in the region far from the core.
Two neutrons are not bound in a free space; 
thus, dineutron correlation vanishes in the asymptotic region far from the core, 
meaning that the dineutron size becomes infinitely large 
as the distance from the core increases, as discussed in Ref.~\cite{hagino07}. 
We call such a breaking effect at the region far from the core 
``dispersion'' of a dineutron. 

To clarify the properties of the dineutron correlation in neutron-rich nuclei, 
we use the extended method of antisymmetrized molecular dynamics (AMD) 
\cite{enyo95_LiBe, enyo95_B, enyo01}. 
The AMD framework can describe various structures 
such as deformation and clustering in general nuclei, 
and it is suitable to describe various core structures. 
We extend the AMD framework 
to investigate the degree of dineutron formation at the surface 
and the degree of expansion of a dineutron tail at the farther region 
while taking into account the above-mentioned breaking effects of a dineutron 
based on the concept of a core$+2n$. 
In addition, to analyze the detailed two-neutron motion, 
we propose a method that enables us to visualize the two-neutron spatial distribution around the core. 
As the first step, we apply the extended AMD framework and the method of analysis to $^{10}$Be, 
which has a well-deformed $^8$Be ($2\alpha$) core, 
and discuss the effect of core structure change on the dineutron correlation in this study. 
We show that the core structure significantly affects the dineutron formation 
and distribution around the core. 

This paper is organized as follows. 
In Sec.~\ref{sec:framework}, we explain the framework used 
to describe and analyze the dineutron correlation in neutron-rich nuclei. 
In Sec.~\ref{sec:result}, we apply the framework to the $^{10}$Be system 
and discuss the dineutron formation and distribution around the $^8$Be core.
We summarize our work in Sec.~\ref{sec:summary}.

\section{Framework}
\label{sec:framework}

We consider a total $A$-nucleon system composed of an $(A-2)$-nucleon core and two valence neutrons 
and investigate the two-neutron motion around the core. 
We describe our framework in this section. 

\subsection{AMD wave function}
\label{sec:AMD_wf}

First, we explain the AMD wave function. 
An AMD wave function for an $A$-nucleon system 
is given by the Slater determinant of $A$ single-particle wave functions;
\begin{equation}
\Phi_{\rm AMD}(\boldsymbol{Z}) = 
\frac{1}{\sqrt{A!}} \mathcal{A} \left\{ \varphi_1 \cdots \varphi_{A} \right\}.
\label{eq:AMD_wf}
\end{equation}
Here $\mathcal{A}$ is an antisymmetrizer
and $\varphi_i \ (i=1,\ldots, A)$ are the single-particle wave functions 
composed of the Gaussian-type spatial part $\phi_i$, 
the spin part $\chi_i$, and the isospin part $\tau_i$ as follows: 
\begin{align}
\varphi_i = & \phi_i
\chi_i \tau_i, 
\label{eq:sp_wf} \\
& \phi_i ( \boldsymbol{r}_j) 
= \left( \frac{2\nu}{\pi} \right)^{3/4} \exp 
\left[ -\nu ( \boldsymbol{r}_j - \boldsymbol{Y}_i)^2 \right], 
\label{eq:sp_wf_space} \\
& \chi_i = \xi_{i \uparrow}\chi_{\uparrow} + \xi_{i \downarrow}\chi_{\downarrow}, 
\label{eq:sp_wf_spin} \\
& \tau_i = p \ {\rm or} \ n. 
\label{eq:sp_wf_isospin}
\end{align}
$\boldsymbol{Z} \equiv \left\{ \boldsymbol{Y}_1, \ldots, \boldsymbol{Y}_A, 
\boldsymbol{\xi}_1, \ldots, \boldsymbol{\xi}_A \right\}$ in Eq.~(\ref{eq:AMD_wf})
are the variational parameters 
that characterize the Gaussian centers and spin orientations 
of the $A$ nucleons. 
$\nu$ is the Gaussian width characterizing the size of the single-nucleon motion,
which is generally common for all nucleons in the standard AMD framework. 

In the present AMD framework, 
the widths can differ from one another 
($\nu_i$ also has the label of the $i$th single-particle state)
and we choose $\nu_i = \nu_c \ (i \leq A-2 {\rm; core})$ 
and $\nu_i = \nu_{2n} \ (i \geq A-1 {\rm; two \ valence \ neutrons} )$, 
as in Ref.~\cite{furutachi09}.
Note that in the case where all the Gaussian widths differ, 
the center of mass motion cannot be separated exactly. 
Therefore, we treat the center of mass motion approximately 
by reducing the center of mass kinetic energy from the total energy, 
as described in \ref{sec:hamiltonian}. 

We comment upon the relation between the present AMD framework
and the similar fermionic molecular dynamics (FMD) framework 
\cite{feldmeier95}. 
Our AMD framework, in which all the Gaussian widths are not identical, 
is indeed quite similar to the FMD framework; 
however, we use phenomenological effective interactions in the Hamiltonian (Eq.~(\ref{eq:hamiltonian})), 
whereas, in a recent FMD calculation, they used effective interactions 
derived from realistic interactions via the unitary correlation operator method \cite{neff03}.

\subsection{Core$+2n$ AMD wave function}
\label{sec:d2n_AMD}

We aim to efficiently describe dineutron correlation in neutron-rich nuclei. 
To this end, we prepare the total wave function in two steps. 
The $A$-nucleon system is regarded as a core composed of $A-2$ nucleons and two valence neutrons. 
The two-step treatment of this system is explained below.  

First, we prepare the core wave function composed of $A-2$ nucleons. 
\begin{equation}
\Phi_{\rm core}^k = \mathcal{A}\left\{ \varphi_1^k, \ldots, \varphi_{A-2}^k \right\}.
\label{eq:core_wf}
\end{equation}
$k$ is the label characterizing the core structure, 
and the core wave function is specified by the parameter sets 
$\{ \boldsymbol{Y}_1^k, \ldots, \boldsymbol{Y}_{A-2}^k, 
\boldsymbol{\xi}_1^k, \ldots, \boldsymbol{\xi}_{A-2}^k \}$. 
We superpose the core wave functions ($\sum_k$) 
to consider the core structure change. 
Here, we consider the $^8$Be core, as shown later. 
We use the 2$\alpha$ cluster wave function for the $^8$Be core for simplicity 
and choose the $\alpha$-$\alpha$ distance $d_{\alpha}$ 
to parameterize the core structure. 
We can simply generalize the core wave function to the AMD wave function. 
For example, in principle, it is possible to
prepare the deformed cores by performing the $\beta$-constraint variation, 
and the deformation parameter $\beta$ is chosen as the parameter for the core structure.   

After preparing the basis wave functions of the core, 
we express the $A$-nucleon total wave functions 
by distributing two neutrons around each core wave function: 
\begin{equation}
\Phi_{\rm total}^k = \frac{1}{\sqrt{A!}} \mathcal{A} \left\{ \Phi_{\rm core}^k \
\varphi_{A-1} \varphi_{A} \right\}, 
\label{eq:total_wf}
\end{equation}
with $\Phi_{\rm total}^k$ being an AMD wave function having the parameters 
$\{ \boldsymbol{Y}_1^k, \ldots, \boldsymbol{Y}_{A-2}^k, 
\boldsymbol{Y}_{A-1}, \boldsymbol{Y}_A,
\boldsymbol{\xi}_1^k, \ldots, \boldsymbol{\xi}_{A-2}^k, 
\boldsymbol{\xi}_{A-1}, \boldsymbol{\xi}_A \}$, 
where the parameters for the core are fixed. 
We perform energy variation only on the parameters 
$\left\{ \boldsymbol{Y}_{A-1}, \boldsymbol{Y}_A, \boldsymbol{\xi}_{A-1}, \boldsymbol{\xi}_A \right\}$ 
for the two valence neutrons 
to minimize the total energy under the condition 
$\sum_i \boldsymbol{Y}_i = (0,0,0)$ with 
$\boldsymbol{Y}_i = \boldsymbol{Y}_i^k - (\boldsymbol{Y}_{A-1}+\boldsymbol{Y}_A)/2$ for $i \leq A-2$. 
The energy variation for two valence neutrons
is performed under the constraint on the distance of the center of mass, 
$(\boldsymbol{Y}_{A-1}+\boldsymbol{Y}_A)/2$, of two neutrons from the core, 
which we call $d_{2n}$: 
\begin{equation}
d_{2n} \equiv \left| \frac{1}{2} \left( \boldsymbol{Y}_{A-1}+\boldsymbol{Y}_A \right)
- \frac{1}{A-2} \sum_{i \leq A-2} \boldsymbol{Y}_i \right|.
\label{eq:d2n_constraint}
\end{equation}
Performing the $d_{2n}$-constraint energy variation for $d_{2n} = d_{2n,m} \ (m=1,\ldots)$, 
we calculate the two-neutron wave functions near and far from the core. 
The variations of $\boldsymbol{Y}_{A-1}$ and $\boldsymbol{Y}_A$ are performed independently 
so the two neutrons near the core tend to have the opposite momentum 
(i.e. the imaginary parts are opposite to one another) 
and tend to be broken by the $LS$ dissociation effect due to the spin-orbit potential from the core. 
On the other hand, two neutrons far from the core tend to have almost the same position, 
indicating the expansion of a spin-singlet dineutron far from the core 
owing to the energy variation. 
A similar constraint is proposed for $\alpha$ cluster in Ref.~\cite{taniguchi04} 
and can efficiently describe $\alpha$ cluster development. 

For the valence neutrons, 
we choose different values of the Gaussian width from those of the core nucleons
($\nu_{2n} \neq \nu_c$). 
Using the different values of the Gaussian widths $\nu_{2n}$,  
the dineutron dispersion effect is taken into account. 
Moreover, the description of the spatially expanded tail parts 
of single-particle wave functions of valence neutrons 
such as the neutron-halo tail should be improved
using multi-range Gaussians. 

We describe the total system 
by superposing all wave functions with core deformation ($k$), 
core-$2n$ distance ($d_{2n}$) and 2n size ($\nu_{2n}$): 
\begin{equation}
\Phi_{A(J\pi)} = \sum_K \sum_k \sum_{d_{2n}} \sum_{\nu_{2n}}
c_{K \kappa} \mathcal{P}^{J \pi}_{MK} 
\Phi_{\rm total}^k(\boldsymbol{Z},\nu_{2n}).
\label{eq:A_wf}
\end{equation}
$\mathcal{P}^{J \pi}_{MK}$ is the parity and angular momentum projection operator to 
the eigenstate of $J \pi$. 
The coefficients $c_{K \kappa}$ 
($\kappa$ is the abbreviation of $\{ k, d_{2n}, \nu_{2n} \}$)
are determined by diagonalizing the Hamiltonian 
(Eq.~(\ref{eq:hamiltonian})). 
Superposing the wave functions with the set of $\{ k, d_{2n}, \nu_{2n} \}$, 
we can take the core structure change into account
as well as the two-neutron motion near and far from the core
while varying their size.

\subsection{$^{10}$Be AMD wave function}
\label{sec:10Be_wf}

Here, we specify the form of the AMD wave function for $^{10}$Be. 
The explicit form of the $^{10}$Be wave function is given by  
\begin{equation}
\Phi_{^{10}{\rm Be}(0+)} = \sum_K \sum_{d_{\alpha}} \sum_{d_{2n}} \sum_{\nu_{2n}}
c_{K \kappa} \mathcal{P}^{0+}_{00} 
\Phi_{\rm total}^{d_{\alpha}} (\boldsymbol{Z}, \nu_{2n}).
\label{eq:10Be_wf}
\end{equation}
$^{10}$Be is regarded as a $^8$Be core with two valence neutrons. 

The $^8$Be core is described with the $2\alpha$ cluster wave function for simplicity.
An $\alpha$ cluster is composed of four spin- and isospin-saturated nucleons in $(0s)^4$
with the same Gaussian center and width. 
We choose $\alpha$-$\alpha$ distance $d_{\alpha}$ 
as the parameter characterizing the core structure
($k=d_{\alpha}$ in Eq.~(\ref{eq:A_wf})). 
We prepare three types of cores with $d_{\alpha} = 2, \ 3, \ 4$ fm
to describe the core structure fluctuation. 
The Gaussian widths of the core nucleons are fixed to $\nu_c = 0.235 \ {\rm fm}^{-2}$, 
giving the lowest energy of an $\alpha$.

Then, we distribute two neutrons around each core with $d_{\alpha} = 2, \ 3, \ 4$ fm 
under the $d_{2n}$-constraint (Eq.~(\ref{eq:d2n_constraint})). 
Here, we choose $d_{2n}$ to be
$d_{2n} = d_{\rm min}d^{m-1} \ (m=1,\ldots,m_{\rm max})$. 
We define $d_{\rm min}d^{m_{\rm max}-1}=d_{\rm max}$. 
We set $(m_{\rm max},d_{\rm min}, d_{\rm max}) = (8,1,5)$.  
We verified that the $^{10}$Be ground state energy is lowered by only 100 keV at most
if the number of bases or the maximum value of $d_{\rm max}$ is increased. 
When $d_{2n}$ is small and two neutrons are distributed near the core, 
the $LS$ dissociation for a dineutron plays an important role 
and the two neutrons mainly occupy the lowest allowed orbit $0p_{3/2}$ 
to gain the spin-orbit potential from the core. 
On the other hand, when $d_{2n}$ is large, 
the tail of the dineutron center of mass motion far from the core can be well described. 

In each $d_{2n}$-constraint variation, 
we fix the Gaussian widths of the two valence neutrons to be
$\nu_{2n} = 0.235, \ 0.125, \ {\rm or} \ 0.08 \ {\rm fm}^{-2}$ 
and perform variation for each $\nu_{2n}$ value. 
The superposition of three types of bases with different $\nu_{2n}$ values for each $d_{2n}$ 
can describe the dineutron size change. 

At the nuclear surface, the $LS$ dissociation of a dineutron is predominant and dineutron correlation is suppressed. 
To more effectively describe the dineutron formation at the surface, 
we additionally superpose the bases prepared with the $d_{2n}$-constraint variation 
without the spin-orbit force ($v_{LS} = 0$ MeV) when $d_{2n}$ is small. 
Here, we prepare the bases without the spin-orbit force when $d_{2n} < 2$ fm 
(four bases) for each $(d_{\alpha},\nu_{2n})$ set. 

We summarize the number of bases used to describe $^{10}$Be. 
We use three types of $^8$Be cores ($d_{\alpha}=2, \ 3, \ 4$ fm) 
and three widths of the two valence neutrons for each core
($\nu_{2n}=0.235, \ 0.125, \ 0.08 {\rm fm}^{-2}$). 
We perform the energy variation on the two valence neutrons 
in each $(d_{\alpha},\nu_{2n})$ set under the $d_{2n}$ constraint. 
We choose eight values for $d_{2n}$ with $v_{LS}=1600$ MeV 
(this value will be explained later)
and four values for $d_{2n}(<2 \ {\rm fm})$ with $v_{LS} = 0$ MeV. 
Then, the number of bases used to describe $^{10}$Be in the present full calculation 
is $(8+4) \times 3 \times 3 = 108$. 

In this study, we would like to examine the effect of core structure 
on dineutron correlation in $^{10}$Be. 
To this end, we compare the results obtained by two types of calculations. 
One is the full calculation mentioned above 
where the bases with $d_{\alpha}=2, \ 3, \ 4$ fm are superposed 
to describe the fluctuation of the $\alpha$-$\alpha$ distance in the core structure, 
i.e., the shape fluctuation of the core. 
We denote the full calculation as the ``full-$d_{\alpha}$'' calculation. 
The other is the truncated calculation where only the bases with $d_{\alpha}=2$ fm are superposed, 
which gives the state with the smaller core deformation. 
We denote this calculation as the ``fixed-$d_{\alpha}$'' calculation. 
The bases in the fixed-$d_{\alpha}$ calculation are recalculated 
because we use the interaction parameters 
modified from those of the full-$d_{\alpha}$ calculation  
to reproduce the two-neutron separation energy in each calculation.

\subsection{$2n$ overlap function}
\label{sec:2n_overlap}
For analysis to examine the dineutron correlation around the core,  
we propose a method using a $2n$ overlap function.
The $2n$ overlap function $f$ is defined as  
\begin{align}
f(\boldsymbol{r}, \boldsymbol{r}_G) \equiv
\mathcal{A} \langle \Phi_{\rm core}^{d_{\alpha}} & 
\delta(\boldsymbol{r}_{A-1} - \boldsymbol{r}_{n_1})
\chi_{\uparrow} \tau_n 
\nonumber \\
\times & \delta(\boldsymbol{r}_{A} - \boldsymbol{r}_{n_2})
\chi_{\downarrow} \tau_n \ | \Phi_{^{10}{\rm Be}(0+)} \rangle,
\label{eq:2n_overlap} \\
\boldsymbol{r} = \boldsymbol{r}_{n_2}-\boldsymbol{r}_{n_1}, & \
\boldsymbol{r}_G = (\boldsymbol{r}_{n_1}+\boldsymbol{r}_{n_2})/2, 
\label{eq:r_coordinate}
\end{align}
where $\Phi_{\rm core}^{d_{\alpha}}$ is the $^8$Be core wave function (Eq.~(\ref{eq:core_wf})) 
specified by the parameter $d_{\alpha}$. 
We use the core with $d_{\alpha}=3$ fm in the full-$d_{\alpha}$ calculation 
and that with $d_{\alpha} = 2$ fm in the fixed-$d_{\alpha}$ calculation, 
which gives the largest overlap with the ground state obtained in each calculation. 
$f(\boldsymbol{r}, \boldsymbol{r}_G)$ is defined as a function of 
the relative and center of mass coordinates of two neutrons
defined by the two-neutron coordinates, 
$\boldsymbol{r}_{n_1}$ and $\boldsymbol{r}_{n_2}$, 
as shown in Eq.~(\ref{eq:r_coordinate}). 
We omit the recoil of the core 
and locate its center of mass in the bra and ket states at the origin 
to measure the largest overlap between these states. 
The quantity $f(\boldsymbol{r}, \boldsymbol{r}_G)$ brings out information about the spatial distribution 
of spin-up and -down neutrons in $\Phi_{^{10}{\rm Be}}$ 
as a function of $\boldsymbol{r}$ and $\boldsymbol{r}_G$. 
This quantity corresponds to a type of reduced width amplitude extended to the three-body case. 
Although the definition of the two-body density is nontrivial, 
we define $f(\boldsymbol{r},\boldsymbol{r}_G)$ to be totally antisymmetrized, 
and $f(\boldsymbol{r},\boldsymbol{r}_G)$ directly reflects the information 
about the spatial distribution of the two neutrons. 
The $2n$ overlap function is useful for analysis of the dineutron correlation, 
as accomplished using the three-body model. 

In particular, we are interested in the component of the dineutron where two neutrons 
are coupled to a spin singlet with a relative $s$ wave ($l=0$).  
We focus on the dineutron in the $S$ wave ($L=0$) with respect to the core for simplicity. 
Here, we use the label ``s'' ($l=0$) for the angular momentum for the relative coordinate $\boldsymbol{r}$  
and ``S'' ($L=0$) for the angular momentum for the center of mass coordinate $\boldsymbol{r}_G$. 
Thus, we project the angular momenta of the relative and center of mass motions of the two neutrons 
to the $s$ and $S$ waves, respectively, 
and consider the $2n$ overlap function, $f^{S=0}_{l=L=0}(r,r_G)$, 
for the states projected to $l=L=0$ as a function of $r=|\boldsymbol{r}|$ and $r_G=|\boldsymbol{r}_G|$. 
The details of the calculation of $f^{S=0}_{l=L=0}$ are given in the Appendix. 

Using the $2n$ overlap function $f^{S=0}_{l=L=0}$, 
we calculate the probability  
\begin{equation}
P^{S=0}_{l=L=0} \equiv \int r^2 dr \ r_G^2 dr_G \ 
\left|f^{S=0}_{l=L=0}(r,r_G) \right|^2,
\label{eq:P_S=0}
\end{equation} 
which corresponds to the component of the spin-singlet and relative $s$-wave 2n pair 
moving in the $S$ wave around the core.

We also calculate the root-mean-square distance for $r$ and $r_G$ of the two neutrons 
in the $S=0$ and $l=L=0$ components:
\begin{align}
& \sqrt{\langle r^2 \rangle^{S=0}_{l=L=0}} 
\nonumber \\ 
& = \left( \int r^2 dr \ r_G^2 dr_G \ r^2 \left| f^{S=0}_{l=L=0}(r,r_G) \right|^2 / P^{S=0}_{l=L=0} \right)^{1/2}, 
\label{eq:r_2n} \\
& \sqrt{\langle r_G^2 \rangle^{S=0}_{l=L=0}} 
\nonumber \\
& = \left( \int r^2 dr \ r_G^2 dr_G \ r_G^2 \left| f^{S=0}_{l=L=0}(r,r_G) \right|^2 / P^{S=0}_{l=L=0} \right)^{1/2}, 
\label{eq:r_G}
\end{align}
which correspond to the dineutron size and the expansion from the core, respectively.

Note that the asymmetry with respect to the exchange, $r/2 \leftrightarrow r_G$ 
in $f^{S=0}_{l=L=0}(r,r_G)$, 
reflects the mixing of single-particle orbits for two neutrons of different parity 
in the $S=0$ and $l=L=0$ components. 
We decompose $f^{S=0}_{l=L=0}$ and $P^{S=0}_{l=L=0}$ 
into symmetric and antisymmetric components as follows: 
\begin{align}
& f^{S=0}_{l=L=0}(r,r_G) 
\nonumber \\
& \hspace{1em}
= f^{S=0(++)}_{l=L=0}(r,r_G) 
+ f^{S=0(--)}_{l=L=0}(r,r_G), 
\label{eq:f_sum} \\
& P^{S=0}_{l=L=0} = P^{S=0(++)}_{l=L=0} + P^{S=0(--)}_{l=L=0}, 
\label{eq:P_sum}
\end{align}
where $f^{S=0(\pm \pm)}_{l=L=0}$ and $P^{S=0(\pm \pm)}_{l=L=0}$ are defined as 
\begin{align}
f^{S=0(\pm \pm)}_{l=L=0}(r,r_G) 
& \equiv \frac{1 \pm \mathcal{P}_{r/2 \leftrightarrow r_G}}{2} f^{S=0}_{l=L=0}(r,r_G) 
\nonumber \\
& = \frac{1}{2} \left(
f^{S=0}_{l=L=0}(r,r_G) \pm f^{S=0}_{l=L=0}(2r_G,r/2) \right), 
\label{eq:f_parity} \\
P^{S=0(\pm \pm)}_{l=L=0} & \equiv \int r^2 dr \ r_G^2 dr_G \ 
\left| f^{S=0(\pm \pm)}_{l=L=0}(r,r_G) \right|^2. 
\label{eq:P_parity}
\end{align}
Here $\mathcal{P}_{r/2 \leftrightarrow r_G}$ is the $r/2 \leftrightarrow r_G$ exchange operator  
and the projection $(1 \pm \mathcal{P}_{r/2 \leftrightarrow r_G})/2$ is equivalent to 
the double projection $(1 \pm \mathcal{P}_{\boldsymbol{r}_{n1}})/2 \times 
(1 \pm \mathcal{P}_{\boldsymbol{r}_{n2}})/2$ 
of the single-particle parities on $f^{S=0}_{l=L=0}(r/2,r_G)$, as shown in the Appendix.  
$\mathcal{P}_{\boldsymbol{r}_{n{1,2}}}$ are the space reflection operators 
of $\boldsymbol{r}_{n_{1,2}} \rightarrow -\boldsymbol{r}_{n_{1,2}}$. 
This means that $f^{S=0(++)}_{l=L=0}$ ($f^{S=0(--)}_{l=L=0}$) 
and $P^{S=0(++)}_{l=L=0}$ ($P^{S=0(--)}_{l=L=0}$) 
indicate the contributions of pure positive- (negative-) parity single-particle states of two neutrons 
in $f^{S=0}_{l=L=0}$ and $P^{S=0}_{l=L=0}$. 
We hereafter label $f^{S=0(\pm \pm)}_{l=L=0}$ as 
the $(++)$ or $(--)$ component of the $2n$ overlap function. 
The dineutron correlation in $^{10}$Be is seen in $f^{S=0}_{l=L=0}$ as 
the coherent mixing of the minor $f^{S=0(++)}_{l=L=0}$ 
into the major $f^{S=0(--)}_{l=L=0}$, 
as shown in Sec.~\ref{sec:10Be_dineutron}.

\section{Result}
\label{sec:result}

\subsection{Effective Hamiltonian}
\label{sec:hamiltonian}

In the present work, we use the Hamiltonian
\begin{equation}
H = T - T_G + V_{\rm cent} + V_{\rm LS} + V_{\rm Coul},
\label{eq:hamiltonian}
\end{equation}
where $T$ and $T_G$ are the total and center of mass kinetic energies. 
In the present framework, since all the Gaussian widths in the present AMD wave functions are not equal, 
the center of mass motion cannot be removed exactly. 
We therefore treat the center of mass motion approximately 
by reducing the expectation value of $T_G$ from the total Hamiltonian. 
$V_{\rm Coul}$ is the Coulomb force that is approximated by the summation of seven Gaussians. 
$V_{\rm cent}$ and $V_{\rm LS}$ are the effective central and spin-orbit interactions. 
We use the Volkov No.2 force \cite{volkov65} as $V_{\rm cent}$
and the spin-orbit part of the G3RS force \cite{tamagaki68} as $V_{\rm LS}$. 
In this work, we choose the strength of the spin-orbit force 
to be $v_{\rm LS} = 1600$ MeV, 
as has been used in the previous works on the subject of $^{10}$Be \cite{suhara10,kobayashi11}. 
The Bartlett and Heisenberg parameters in the central force are $b=h=0.125$ 
which reproduce the deuteron binding energy and the $n$-$n$ unbound feature. 
We use the Majorana parameter $m=0.60$ as it was used in Refs.~\cite{suhara10,kobayashi11} 
in the full-$d_{\alpha}$ calculation. 
In the fixed-$d_{\alpha}$ calculation, 
we choose $m=0.64$ to give almost the same two-neutron separation energy, $S_{2n}$, 
that was obtained in the full-$d_{\alpha}$ calculation.

\subsection{$^{10}$Be fundamental properties}
\label{sec:10Be_property}

\begin{table}
\caption{The used Majorana parameter $m$, the two neutron separation energy $S_{2n}$, 
the root-mean-square radii $r_{m,p,n}$ of matter, protons and neutrons, 
and the expectation value of the squared neutron total spin $\langle S_n^2 \rangle$ 
in the full-$d_{\alpha}$ (full) and fixed-$d_{\alpha}$ (fixed) calculations.
The experimental values of the matter radii are referred from Ref.~\cite{ozawa01}.}
\label{tab:10Be_property}
\begin{tabular}{cccccccc}
\hline \hline
&& $m$ & $S_{2n}$ (MeV) & $r_m$ (fm) & $r_p$ (fm) & $r_n$ (fm) & $\langle S_n^2 \rangle$ \\ 
\hline
full && $0.60$ & $6.71$ & $2.43$ & $2.22$ & $2.51$ & $0.39$ \\
fixed && $0.64$ & $6.34$ & $2.33$ & $2.01$ & $2.38$ & $0.50$ \\
Expt. && & $8.48$ & $2.30 \pm 0.02$ & \\
\hline
\hline
\end{tabular}
\end{table}

We calculate the two-neutron separation energy, $S_{2n}$, 
the matter, proton and neutron radii, $r_{m,p,n}$, 
and the expectation value of the squared neutron total spin, $\langle S_n^2 \rangle$, 
shown in Table~\ref{tab:10Be_property}. 
The two-neutron separation energy is calculated as
the difference between the total binding energy and the core binding energy, 
\begin{align}
S_{2n} = - \Big( & \langle \Phi_{^{10}{\rm Be}(0+)}|H| \Phi_{^{10}{\rm Be(0+)}} \rangle 
\nonumber \\
& - \langle \Phi_{^8 {\rm Be(0+)}}|H| \Phi_{^8 {\rm Be(0+)}} \rangle \Big), 
\label{eq:S2n}
\end{align}
where $\Phi_{^8 {\rm Be(0+)}}$ is the superposition 
of the $^8$Be core wave functions with $d_{\alpha}=2, \ 3, \ 4$ fm 
in the full-$d_{\alpha}$ calculation 
and that with $d_{\alpha}=2$ fm in the fixed-$d_{\alpha}$ calculation, 
and they are projected onto $J^{\pi}=0^+$. 
The binding energy of $^{10}$Be in each calculation is 
$-60.42$ MeV (full-$d_{\alpha}$) 
and $-49.63$ MeV (fixed-$d_{\alpha}$).  
The root-mean-square radii of matter, protons and neutrons are larger in the full-$d_{\alpha}$ calculation 
than those in the fixed-$d_{\alpha}$ calculation. 
This is natural because the core size becomes larger in the full-$d_{\alpha}$ calculation 
due to the fluctuation in the $\alpha$-$\alpha$ distance.

We show the neutron spin expectation value, $\langle S_n^2 \rangle$, in Table~\ref{tab:10Be_property}. 
In the present calculation, 
the $^8$Be core has zero proton- and neutron-spins 
and, therefore, $\langle S_n^2 \rangle$ indicates the squared spin expectation value
of the two valence neutrons. 
The finite value of $\langle S_n^2 \rangle$ reflects the spin-triplet component; 
in other words, the degree of the $LS$ dissociation of a dineutron. 
$\langle S_n^2 \rangle$ is larger in the fixed-$d_{\alpha}$ calculation than that in the full-$d_{\alpha}$ calculation, 
indicating that the $LS$ dissociation increases 
when the core structure is fixed to be small. 
This point is discussed later in connection with the dineutron enhancement 
due to the core structure change. 

We have checked the relationship between the present AMD wave function 
and the DC wave function used in our previous work \cite{kobayashi11}. 
For the DC wave function, we assume a spin-singlet $2n$ pair around a core, 
and we have superposed the DC wave functions with the AMD wave functions. 
The main role of the DC wave function discussed in Ref.~\cite{kobayashi11} 
is to describe the dineutron-tail component in $^{10}$Be. 
If we superpose the DC wave functions used in Ref.~\cite{kobayashi11} 
with the present AMD wave functions, 
the ground state energy is lowered by only 300 keV at most
and the other properties are largely unchanged. 
This means that the contribution of the DC wave function is minor, 
and that the structure as well as the dineutron tail can be well-described 
with the present $d_{2n}$-constrained AMD wave functions, 
at least for a $^{10}$Be system with sufficiently bound valence neutrons.

\subsection{Effect of core structure change on dineutron correlation in $^{10}$Be}
\label{sec:10Be_dineutron}

In this section, we discuss the dineutron correlation in the ground state of $^{10}$Be, 
focusing mainly on the effect of $^8$Be core structure change. 

\begin{figure}[t!]
\includegraphics[scale=0.8]{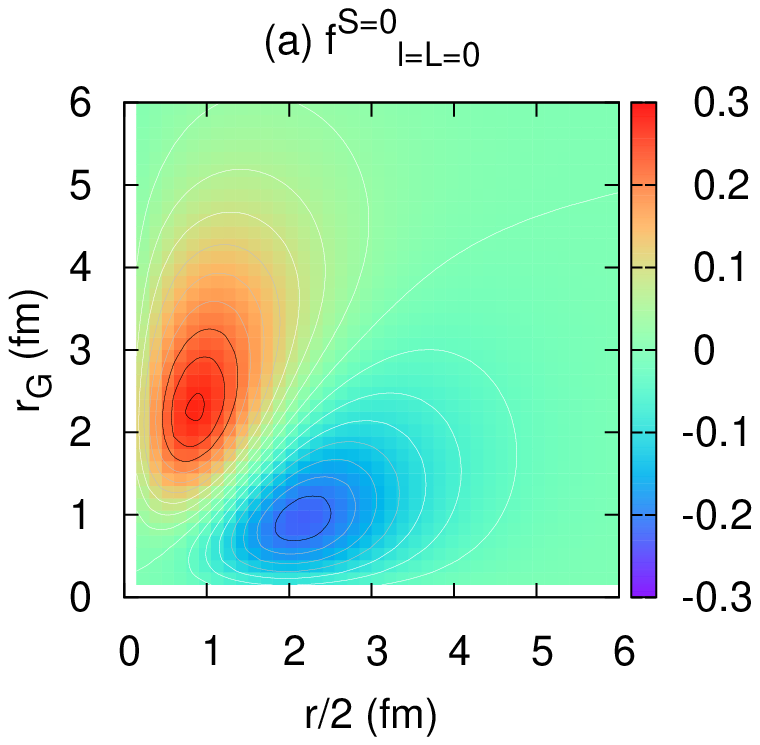} \\
\includegraphics[scale=0.8]{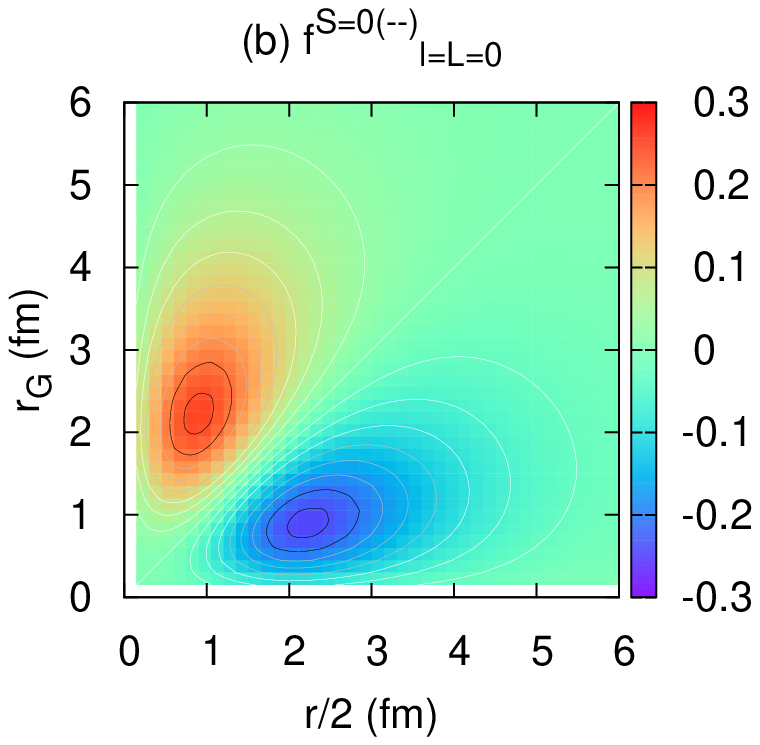} \\
\includegraphics[scale=0.8]{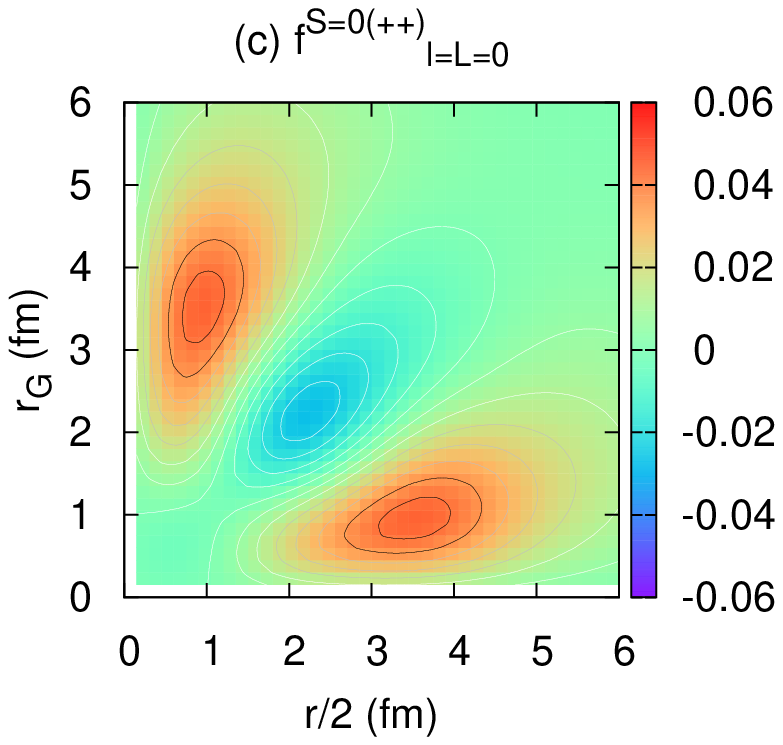}
\caption{(Color online) The $2n$ overlap functions of (a)$f^{S=0}_{l=L=0}$, 
(b)$f^{S=0(--)}_{l=L=0}$ and (c)$f^{S=0(++)}_{l=L=0}$ 
in the full-$d_{\alpha}$ calculation as functions of $(r/2, r_G)$. 
For the guide, the colors of the lines corresponding to the region
where the absolute amplitude is largest, intermediate and smallest
are black, gray, and white, respectively. }
\label{fig:2n_overlap}
\end{figure}

As an example of the analysis of dineutron correlation using the $2n$ overlap function, 
we plot the $2n$ overlap functions, $f^{S=0}_{l=L=0}$ and $f^{S=0(\pm \pm)}_{l=L=0}$, 
in the full-$d_{\alpha}$ calculation 
as functions of $(r/2,r_G)$ in Fig.~\ref{fig:2n_overlap}. 
To show the asymmetry with respect to the $r_G = r/2$ line, 
we show the $r/2$-$r_G$ plot instead of the $r$-$r_G$ plot. 
There are two peaks in $f^{S=0}_{l=L=0}$ and $f^{S=0(--)}_{l=L=0}$ 
and three peaks in $f^{S=0(++)}_{l=L=0}$. 
Hereafter, we refer to the peak in the $r_G > r/2$ region of $f^{S=0}_{l=L=0}$ as the dineutron peak 
and the peak in the $r_G < r/2$ region of $f^{S=0}_{l=L=0}$ as the cigar peak. 
The two-peak structure in $f^{S=0(--)}_{l=L=0}$ comes from the two neutrons occupying the $p^2$ orbits
and the three-peak structure in $f^{S=0(++)}_{l=L=0}$ comes from those occupying the $(sd)^2$ orbits.
The $(--)$ and $(++)$ components of the $2n$ overlap function, 
$f^{S=0(--)}_{l=L=0}$ and $f^{S=0(++)}_{l=L=0}$, 
are antisymmetric and symmetric with respect to the $r_G=r/2$ line, respectively. 
In general, without the mixing of different parity contributions for the single-particle orbits, 
the absolute amplitudes of the dineutron and cigar peaks are exactly the same, 
and we do not describe the case as dineutron correlation. 
In other words, 
dineutron correlation is reflected in the asymmetry 
between the dineutron and cigar peaks with respect to the $r_G = r/2$ line in $f^{S=0}_{l=L=0}$. 
In the present case, the asymmetry in $f^{S=0}_{l=L=0}$ comes from the mixing of 
the minor $(sd)^2$ component ($f^{S=0(++)}_{l=L=0}$)
into the major $p^2$ component ($f^{S=0(--)}_{l=L=0}$), 
as shown in Eq.~(\ref{eq:f_sum}). 
The relative phases at the dineutron peaks ($r/2 \sim 1$ fm) in $f^{S=0(--)}_{l=L=0}$ 
and $f^{S=0(++)}_{l=L=0}$ are coherent; 
on the other hand, those at the cigar peaks ($r_G \sim 1$ fm) 
in $f^{S=0(--)}_{l=L=0}$ and $f^{S=0(++)}_{l=L=0}$ are incoherent, 
leading to asymmetry between the dineutron and cigar peaks.  
The dineutron enhancement is seen in the obvious asymmetry between the dineutron and cigar components 
coming from significant mixing of the different-parity single-particle orbits for each valence neutron.

\begin{table*}
\caption{The probabilities of the spin-singlet $2n$ pair $P^{S=0}_{l=L=0}$, 
those of the positive- or negative-parity components $P^{S=0(\pm \pm)}_{l=L=0}$, 
and the root-mean-square expectation values of the relative and center of mass distances of the $2n$ pair 
$\sqrt{\langle r^2 \rangle^{S=0}_{l=L=0}}$ and $\sqrt{\langle r_G^2 \rangle^{S=0}_{l=L=0}}$, 
in the full-$d_{\alpha}$ (full) and fixed-$d_{\alpha}$ (fixed) calculations.}
\label{tab:10Be_2n_property}
\begin{ruledtabular}
\begin{tabular}{ccccccc}
 && $P^{S=0}_{l=L=0}$ & $P^{S=0(--)}_{l=L=0}$ & $P^{S=0(++)}_{l=L=0}$ &
$\sqrt{\langle r^2 \rangle^{S=0}_{l=L=0}}$ (fm) & $\sqrt{\langle r_G^2 \rangle^{S=0}_{l=L=0}}$ (fm)\\ 
\hline
full && 0.515 & 0.491 & 0.025 & 3.70 & 2.53 \\
fixed && 0.546 & 0.530 & 0.016 & 3.75 & 2.37 \\
\end{tabular}
\end{ruledtabular}
\end{table*}

We now investigate the dependence of the degree of dineutron enhancement 
on the core structure by comparing the results 
obtained from the full-$d_{\alpha}$ and fixed-$d_{\alpha}$ calculations. 
In Table.~\ref{tab:10Be_2n_property}, 
we show some properties of the $2n$ pair in two calculations. 
As mentioned above, the mixing of the different-parity single-particle orbits for the two neutrons  
reflects the dineutron correlation, 
in other words, the mixing ratio of $P^{S=0(++)}_{l=L=0}$ into $P^{S=0(--)}_{l=L=0}$ 
reflects the strength of dineutron correlation. 
The mixing ratio $P^{S=0(++)}_{l=L=0}/P^{S=0(--)}_{l=L=0}$ is 
$5.1$ \% in the full-$d_{\alpha}$ calculation 
and $3.0$ \% in the fixed-$d_{\alpha}$ calculation, 
meaning that the large deformation and large shape fluctuation in the $^8$Be core structure 
enhance the dineutron correlation. 
The stronger dineutron correlation in the full-$d_{\alpha}$ calculation is also reflected in 
the slightly smaller value of $\sqrt{\langle r^2 \rangle^{S=0}_{l=L=0}}$ 
than that in the fixed-$d_{\alpha}$ calculation. 

The reason for which the core structure change enhances the dineutron correlation 
is as follows; 
when the core deformation is fixed to be small in the fixed-$d_{\alpha}$ calculation, 
two valence neutrons are distributed near the core to a larger extent 
and they feel the stronger spin-orbit potential at the surface. 
In the $^{10}$Be case, 
the $LS$-favored orbit of $0p_{3/2}$ is partially unoccupied 
and the valence neutrons are favored energetically to occupy the $0p_{3/2}$ orbit. 
Under the smaller core deformation, 
the simplest shell-model component (two neutrons occupy only the lowest shell) 
is predominant because of the spin-orbit potential from the core 
and, therefore, dineutron correlation is not greatly enhanced 
because of the $LS$ dissociation effect.  
On the other hand, when the core deformation becomes larger, 
the mean field generated by the core expands to the farther region
so that two valence neutrons can be radially expanded from the core 
to form a spin-singlet compact dineutron. 
As a result, $\sqrt{\langle r^2_G \rangle^{S=0}_{l=L=0}}$ is larger in the full-$d_{\alpha}$ calculation.
Moreover, the spin-orbit potential becomes weaker in the region far from the core,
resulting in the suppression of the $LS$ dissociation effect on the dineutron. 
The suppression of the $LS$ dissociation in the full-$d_{\alpha}$ calculation 
is seen by the smaller spin-triplet component ($\langle S_n^2 \rangle/2$) 
than that in the fixed-$d_{\alpha}$ calculation, as already shown in Table~\ref{tab:10Be_property}. 

However, it should be noted that the $P^{S=0}_{l=L=0}$ value itself 
is smaller in the full-$d_{\alpha}$ calculation 
than that in the fixed-$d_{\alpha}$ calculation. 
This is due to the fact that, if the core deformation becomes larger, 
the spin-singlet $2n$ components in the strong-coupling channels 
between the core and $2n$ (e.g. $(L_{\rm core}=2) \otimes (L=2) = 0$
which is projected out in the calculations of $P^{S=0}_{l=L=0}$ and $P^{S=0 (\pm \pm)}_{l=L=0}$) 
are mixed to a greater extent. 
We emphasize that 
the mixing of the $(++)$ component of the single-particle orbits of $2n$ 
into the dominant $(--)$ component increases, 
reflecting the enhancement of the dineutron correlation.

\begin{figure}
\includegraphics[scale=0.7]{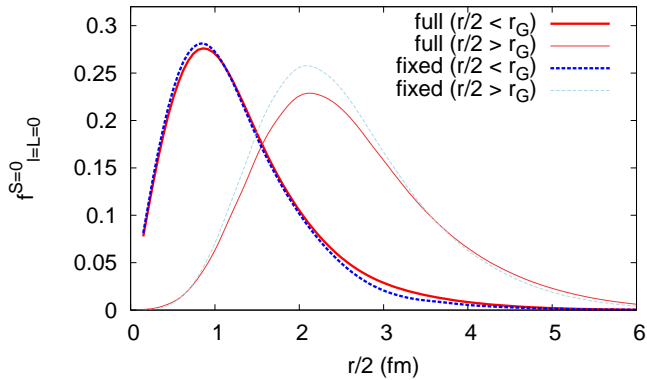} \\
\caption{(Color online) The absolute values of 
the projection of the $2n$ overlap functions of $f^{S=0}_{l=L=0}$ 
onto the $r/2$ axis 
in the full-$d_{\alpha}$ (full) and fixed-$d_{\alpha}$ (fixed) calculations. 
The red solid lines correspond to the full-$d_{\alpha}$ calculation 
and the blue dashed lines correspond to the fixed-$d_{\alpha}$ calculation. 
The thick lines are the dineutron component 
(the amplitudes in the $r/2 < r_G$ region)
and the thin lines are the cigar component
(the amplitudes in the $r/2 > r_G$ region).  }
\label{fig:2n_overlap_r}
\end{figure}

To investigate the effect of the core structure on the $2n$ properties in more detail, 
we compare the $2n$ overlap functions in the full-$d_{\alpha}$ and fixed-$d_{\alpha}$ calculations.
In Fig.~\ref{fig:2n_overlap_r}, 
we plot the absolute values of the $2n$ overlap function, 
$|f^{S=0}_{l=L=0}|$, projected onto the $r/2$ axis 
obtained in the full-$d_{\alpha}$ and fixed-$d_{\alpha}$ calculations. 
It can be seen that 
the dineutron peak ($r/2 \sim 1$ fm) is larger than 
the cigar peak ($r/2 \sim 2$ fm) in both calculations. 
However, the difference between the dineutron and cigar peaks 
is larger in the full-$d_{\alpha}$ calculation 
than that in the fixed-$d_{\alpha}$ calculation. 
This means that the dineutron correlation is enhanced 
due to the fluctuation in the $\alpha$-$\alpha$ distance, as discussed in connection with 
the mixing of $P^{S=0(++)}_{l=L=0}$ into $P^{S=0(--)}_{l=L=0}$. 
It should be noticed that 
the absolute amplitudes of both the dineutron and cigar peaks themselves 
become smaller in the full-$d_{\alpha}$ calculation
because of the mixing of the $(L_{\rm core} \neq 0) \otimes (L \neq 0) = 0$ components, 
as mentioned above.
Introducing the fluctuation in the distance between $2\alpha$s, 
the cigar peak decreases more than the dineutron peak 
and the difference between these peaks is certainly increased, 
indicating dineutron enhancement depending on the core structure change.

\subsection{Utility of the $d_{2n}$-constraint calculation}
\label{sec:d_2n}

\begin{figure}
\includegraphics[scale=0.7]{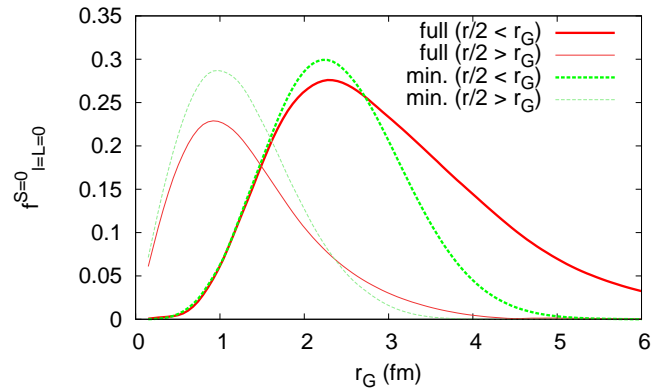}
\caption{(Color online) The absolute values of 
the projection of the $2n$ overlap functions of $f^{S=0}_{l=L=0}$ onto the $r_G$ axis 
in the full-$d_{\alpha}$ calculation 
and the calculation using the basis that gives the minimum energy. 
The red solid lines corresponds to the full-$d_{\alpha}$ (full) calculation 
and the green dashed lines are the calculation using only the minimum-energy basis (min.). 
The thick lines are the dineutron component 
(the amplitudes in the $r/2 < r_G$ region)
and the thin lines are the cigar component
(the amplitudes in the $r/2 > r_G$ region).  }
\label{fig:2n_overlap_rG}
\end{figure}

Finally, to show the utility of the present $d_{2n}$-constraint calculation, 
we compare the dineutron-tail component 
obtained by the full-$d_{\alpha}$ calculation with that obtained by the single basis 
having the parameters $(d_{\alpha},d_{2n},\nu_{2n}) = (3,1.8,0.235)$, 
which is the minimum-energy AMD wave function 
with the assumption of a $2\alpha$ core.
The minimum-energy basis has an overlap of $\sim 82$ \% with the state obtained by the full bases calculation, 
and its energy is $-55.80$ MeV, which is $\sim 5$ MeV higher than 
the ground-state energy in the full-$d_{\alpha}$ calculation. 
This indicates that the superposition of the basis wave functions 
significantly improves the $^{10}$Be wave function. 
We plot in Fig.~\ref{fig:2n_overlap_rG} the absolute values of the $2n$ overlap function $|f^{S=0}_{l=L=0}|$ 
projected onto the $r_G$ axis 
for the states obtained in the full-$d_{\alpha}$ calculation 
and the minimum-energy basis calculation. 
In the calculation using the minimum-energy basis, 
the amplitudes of the dineutron ($r_G \sim 2$ fm) and cigar ($r_G \sim 1$ fm) peaks do not differ greatly, 
because the $0p_{3/2}$ orbit (the lowest shell-model limit) is dominant for $2n$
and the dineutron correlation mostly disappears in this basis. 

A remarkable difference between these calculations is seen in the dineutron-tail component ($r_G \gtrsim 4$ fm). 
The minimum-energy basis does not have the dineutron tail expanded into the farther region 
because in a single basis, this tail has a rapidly dumping Gaussian form. 
On the other hand, in the full-$d_{\alpha}$ calculation 
where many bases with various $d_{2n}$ and $\nu_{2n}$ values are superposed, 
the dineutron tail is improved remarkably. 
This indicates that the present framework well-describes the dineutron tail. 
It is also expected that this method can be useful for extremely loosely bound nuclei 
with neutron-halo or -skin structures. 
Further application to loosely bound nuclei is future work.

\section{Summary}
\label{sec:summary}

In this study, we extended the AMD framework to describe dineutron correlation 
around a core with various structures. 
We first prepared the core wave functions using the AMD method, 
which was useful for describing various structure changes such as deformation and clustering. 
We described the detailed two-neutron motion around the cores 
considering the relative distance between the core and the two neutrons as the degree of freedom, 
and we superposed the basis wave functions with various core-$2n$ distances
to describe the competition between the dineutron formation and the $LS$ dissociation at the nuclear surface 
as well as the dineutron expansion far from the core. 
We additionally changed the Gaussian widths of the two neutrons 
to take into account the dineutron dispersion effect. 
Moreover, we have constructed a $2n$ overlap function 
as the analyzing method for investigating dineutron correlation around a core in detail. 
This method enabled us to visualize the spatial correlation between two neutrons 
and made the discussion clear. 

As a first step, we applied the framework to $^{10}$Be 
and focused on the effect of the $^8$Be core structure change on the dineutron correlation.
In the present work, we assumed a $2\alpha$ cluster structure for the $^8$Be core
and considered the fluctuation in the $\alpha$-$\alpha$ distance as the core structure change. 
Dineutron correlation was seen in the degree of mixing of different-parity single-particle orbits, 
i.e., the mixing of the minor $(sd)^2$ component 
into the major $p^2$ component for two neutrons in the $^{10}$Be case. 
When the core was less deformed, 
the two neutrons were distributed near the core 
and greatly dissociated to the independent $(0p_{3/2})^2$ component 
due to the spin-orbit potential from the core; 
as a result, dineutron correlation was suppressed. 
On the other hand, when the core was well-deformed by taking into account the fluctuation 
in the distance between $2\alpha$s, 
two neutrons could be expanded radially and they were favored to form a dineutron. 
Moreover, at the region far from the core, the spin-orbit potential becomes weaker 
and the $LS$ dissociation effect was suppressed. 
As a result, the dineutron correlation was enhanced at the surface and further regions
due to the core structure change. 

In this work, we have applied the extended methods to $^{10}$Be
and showed that the present framework is useful to describe 
the dineutron component around the well-deformed core. 
Additionally, we have concluded that the core structure significantly affects the dineutron correlation. 
Certainly, $^{10}$Be is not a very loosely bound nucleus 
and the $LS$ dissociation effect on the dineutron plays an important role. 
However, we could see the obvious asymmetry between the dineutron and cigar components
 (Fig.~\ref{fig:2n_overlap}), 
and would like to stress that dineutron correlation can be more or less apparent in most neutron-rich nuclei, 
even in those without an outstanding neutron-halo or -skin structure. 
In the future, we will apply the present framework to various nuclei 
and clarify the universal properties of dineutron correlation, 
e.g., the effect of the core excitation and clustering 
in addition to that of the core deformation on dineutron correlation,  
through the systematic investigation.

\appendix*
\section{Overlap functions $f$ and $f^{(\pm \pm)}$}
\label{appendix}

Here we show the details of the overlap functions $f$ and $f^{(\pm \pm)}$.
We define the general $2N$ overlap function $f$ as below: 
\begin{align}
f(\boldsymbol{r}, \boldsymbol{r}_G) \equiv
\mathcal{A} \langle \Phi_{\rm core} & 
\delta(\boldsymbol{r}_{A-1} - \boldsymbol{r}_{N_1})
\chi_{A-1} \tau_{A-1} 
\nonumber \\
\times & \delta(\boldsymbol{r}_{A} - \boldsymbol{r}_{N_2})
\chi_{A} \tau_A \ | \Phi_{\rm total} \rangle,
\label{eq:overlap} \\
\boldsymbol{r} = \boldsymbol{r}_{N_2}-\boldsymbol{r}_{N_1}, & \
\boldsymbol{r}_G = (\boldsymbol{r}_{N_1}+\boldsymbol{r}_{N_2})/2, 
\label{eq:r_coordinate_N}
\end{align}
where $\Phi_{\rm total}$ is the total wave function with $A$ nucleons 
whose center of mass of the core component is located at the origin  
and $\Phi_{\rm core}$ is the core wave function with $A-2$ nucleons
whose center of mass is located at the origin. 
$\chi_{A-1,A}$ and $\tau_{A-1,A}$ are the spin and isospin wave functions 
of the $(A-1)$th or $A$th nucleon 
and we can choose $\uparrow$ or $\downarrow$ for the spin components 
and $p$ or $n$ for the isospin components 
for each purpose. 
In the present work, we want to investigate the dineutron motion 
that is a spin-singlet pair of two neutrons 
so that $\chi_{A-1} = \uparrow$ and  $\chi_{A} = \downarrow$, and $\tau_{A-1, A} = n$. 
In this appendix, we show only this case
but generalization is simple. 

In the case of a spin-singlet $2n$ pair, Eq.~(\ref{eq:overlap}) can be rewritten as 
\begin{widetext}
\begin{align}
f(\boldsymbol{r}, \boldsymbol{r}_G) 
= \sum_{i,j \in n} &
\left( \frac{4 \nu_i \nu_j}{\pi^2} \right)^{3/4}
\exp \left[ - \nu_i ( \boldsymbol{r}_{n_1} - \boldsymbol{Y}_i )^2 
- \nu_j ( \boldsymbol{r}_{n_2} - \boldsymbol{Y}_j )^2
\right] 
\nonumber \\
\times & \langle \chi_{\uparrow}|\chi_i \rangle 
\langle\chi_{\downarrow}|\chi_j \rangle 
\times \det B^{(i,j)} 
\label{eq:overlap2} \\
= \sum_{i,j \in n} &
\left( \frac{4 \nu_i \nu_j}{\pi^2} \right)^{3/4}
\exp \left[ - (\nu_i+\nu_j)(r_G^2 + r^2/4) - \nu_i Y_i^2 - \nu_j Y_j^2
- ( \nu_i - \nu_j )\boldsymbol{r} \cdot \boldsymbol{r}_G/2 \right] 
\nonumber \\
\times & 4 \pi \sum_l \sum_m j_l \left(- i (\nu_i Y_i - \nu_j Y_j)r \right)
Y_{lm} (\hat{\boldsymbol{r}}) Y_{lm}^*(\hat{\boldsymbol{Y}}_r)
\nonumber \\
\times & 4 \pi \sum_l \sum_m j_l \left(-2 i (\nu_i Y_i + \nu_j Y_j)r_G \right)
Y_{lm} (\hat{\boldsymbol{r}}_G) Y_{lm}^*(\hat{\boldsymbol{Y}}_G)
\nonumber \\
\times & \langle \chi_{\uparrow}|\chi_i \rangle 
\langle\chi_{\downarrow}|\chi_j \rangle 
\times \det B^{(i,j)}, 
\label{eq:overlap3}
\end{align}
\end{widetext}
where $j_l$ are the spherical Bessel functions 
and $Y_{lm}$ are the spherical harmonics. 
$\hat{\boldsymbol{Y}}_{r,G}$ are the polar angles 
of the vectors $\nu_i \boldsymbol{Y}_i - \nu_j \boldsymbol{Y}_j$
or $\nu_i \boldsymbol{Y}_i + \nu_j \boldsymbol{Y}_j$,
respectively.  
$B^{(i,j)}$ is an $(A-2)\times(A-2)$ norm matrix 
composed of $\langle \varphi_{{\rm core}, \kappa}| \ (\kappa=1,\ldots,A-2)$ in $\langle \Phi_{\rm core}|$
and $|\varphi_{{\rm total}, \kappa} \rangle \ (\kappa=1,\ldots,A)$ in $|\Phi_{\rm total} \rangle$ 
except for the $i$th and $j$th single-particle wave functions. 
$f(\boldsymbol{r},\boldsymbol{r}_G)$ is totally antisymmetrized by $\mathcal{A}$ 
and hence $B^{(j,i)} = - B^{(i,j)}$. 
We perform the angular integrals 
$\int d^2 \hat{\boldsymbol{r}} Y_{00}(\hat{\boldsymbol{r}})
\int d^2 \hat{\boldsymbol{r}}_G Y_{00}(\hat{\boldsymbol{r}}_G)$
to project the relative and center of mass motions of $2n$ onto $l=L=0$, 
and we additionally neglect the term proportional to $\boldsymbol{r} \cdot \boldsymbol{r}_G$ 
in the exponential term in Eq.~(\ref{eq:overlap3}),  
resulting in $f^{S=0}_{l=L=0}(r,r_G)$ as 
\begin{widetext}
\begin{align} 
f^{S=0}_{l=L=0}(r,r_G) = \sum_{i,j \in n} &
\left( \frac{4 \nu_i \nu_j}{\pi^2} \right)^{3/4}
\exp \left[ - (\nu_i+\nu_j)(r_G^2 + r^2/4) - \nu_i Y_i^2 - \nu_j Y_j^2 \right] 
\nonumber \\
\times & 4 \pi j_0 \left(- i (\nu_i Y_i - \nu_j Y_j)r \right)
\times j_0 \left(-2 i (\nu_i Y_i + \nu_j Y_j)r_G \right) 
\nonumber \\
\times & \langle \chi_{\uparrow}|\chi_i \rangle 
\langle\chi_{\downarrow}|\chi_j \rangle 
\times \det B^{(i,j)}.
\label{eq:overlap4}
\end{align}
\end{widetext}
This $2n$ overlap function only depends on the absolute values 
of $r=|\boldsymbol{r}|$ and $r_G=\boldsymbol{r}_G$. 
We calculate the probability of the spin-singlet $2n$ pair, $P^{S=0}_{l=L=0}$, 
the root mean square distance between two neutrons  
$\sqrt{\langle r^2 \rangle^{S=0}_{l=L=0}}$, 
and that between the two neutrons and the core 
$\sqrt{\langle r_G^2 \rangle^{S=0}_{l=L=0}}$
simply using the numerical integrals of $r$ and $r_G$ 
(Eqs.~(\ref{eq:P_S=0}), (\ref{eq:r_2n}) and (\ref{eq:r_G})). 

We calculate the $(++)$ and $(--)$ components $f^{S=0(\pm \pm)}_{l=L=0}$, 
from the $2n$ overlap function $f^{S=0}_{l=L=0}$ as Eq.~(\ref{eq:f_parity}). 
Here, we show that 
$f^{S=0(\pm \pm)}_{l=L=0}$, defined in Eq.~(\ref{eq:f_parity}), 
is certainly equivalent to the components 
where both neutrons are projected to positive- or negative-parity single-particle orbits 
if the relative and center of mass motions of $2n$ are projected as $l=L=0$. 
We begin by projecting both two neutrons 
in the $2n$ overlap function $f(\boldsymbol{r},\boldsymbol{r}_G)$ 
to the positive- or negative-parity single-particle orbits. 
Noting the definition of the coordinates $\boldsymbol{r}$ and $\boldsymbol{r}_G$ 
in Eq.~(\ref{eq:r_coordinate_N}), we have
\begin{widetext}
\begin{align}
& ( 1 \pm \mathcal{P}_{\boldsymbol{r}_{n1}})/2 
\times ( 1 \pm \mathcal{P}_{\boldsymbol{r}_{n2}})/2 
\times f(\boldsymbol{r},\boldsymbol{r}_G) 
\nonumber \\
& = \frac{1}{4} 
\left[ \left( 1 + \mathcal{P}_{\boldsymbol{r}_{n1}} \mathcal{P}_{\boldsymbol{r}_{n2}} \right)
\pm \left( \mathcal{P}_{\boldsymbol{r}_{n1}} + \mathcal{P}_{\boldsymbol{r}_{n2}} \right) \right] 
f(\boldsymbol{r},\boldsymbol{r}_G) 
\nonumber \\
& = \frac{1}{4} \left[ \left( f(\boldsymbol{r},\boldsymbol{r}_G)
+ f(-\boldsymbol{r},-\boldsymbol{r}_G) \right) 
\pm \left( f(2\boldsymbol{r}_G,\boldsymbol{r}/2)
+ f(-2\boldsymbol{r}_G,-\boldsymbol{r}/2) \right) \right].
\label{eq:f_parity1}
\end{align}
\end{widetext}
If we project $f$ to $l=L=0$ with the operator $\mathcal{P}_{l=L=0}$, 
$f^{S=0}_{l=L=0}$ becomes just a function of the absolute values of $r$ and $r_G$ 
($\mathcal{P}_{l=L=0} f(\boldsymbol{r}, \boldsymbol{r}_G) 
= f^{S=0}_{l=L=0}(r,r_G)$). 
As a result of the projection to $l=L=0$, 
the former two terms and the latter two terms on the last line of Eq.~(\ref{eq:f_parity1})
give the same contributions, respectively. 
We can therefore rewrite the Eq.~(\ref{eq:f_parity1}) 
under the projection to $l=L=0$ as 
\begin{align}
& \mathcal{P}_{l=L=0} \left[ ( 1 \pm \mathcal{P}_{\boldsymbol{r}_{n1}})/2 
\times ( 1 \pm \mathcal{P}_{\boldsymbol{r}_{n2}})/2 
\times f(\boldsymbol{r},\boldsymbol{r}_G) \right]
\nonumber \\
& = \frac{1}{2} \left( f^{S=0}_{l=L=0} (r,r_G)
\pm f^{S=0}_{l=L=0} (2r_G,r/2) \right)
\nonumber \\
& = f^{S=0(\pm \pm)}_{l=L=0}. 
\end{align}
This means that
$f^{S=0(\pm \pm)}_{l=L=0}$ defined in Eq.~(\ref{eq:f_parity}) 
are certainly the components in the $2n$ overlap function where both two neutrons are projected 
to the positive- or negative-parity single-particle orbits 
if the relative and center of mass motions of  the two neutrons are projected as $l=L=0$.

\begin{acknowledgments}
This work was supported by a Grant-in-Aid for Scientific Research 
from Japan Society for the Promotion of Science (JSPS).
A part of the computational calculations of this work was performed by using the
supercomputers at YITP.
\end{acknowledgments}

\bibliography{reference}

\end{document}